\documentclass[twocolumn,aps,superscriptaddress]{revtex4}
 
\usepackage{amsthm,amssymb,amsmath}
\usepackage{graphicx}
\usepackage{dcolumn}
\usepackage{bm}
\usepackage{color} 

\usepackage{graphicx,graphics}
\usepackage{dcolumn}                
\usepackage{amssymb}

\begin{document}

\title{Universal wave functions structure in mixed systems}

\author{Diego A. Wisniacki}

\affiliation{Departamento de F\'\i sica and IFIBA,
FCEyN, UBA Ciudad Universitaria,
Pabell\'on 1, Ciudad Universitaria, 1428 Buenos Aires, Argentina.
}

\begin{abstract}
When a regular classical system is perturbed, non-linear resonances appear as prescribed by the 
KAM and Poincar\`{e}-Birkhoff theorems.  Manifestations of this classical phenomena 
to the morphologies of quantum wave functions are studied in this letter.   We reveal  
a systematic formation of an universal structure of localized wave functions
in systems with mixed classical dynamics.  
Unperturbed states that live around invariant tori
are mixed when they collide in an avoided crossing if their quantum numbers differ in a multiple to the order of 
the classical resonance. At the avoided crossing eigenstates are localized in
the island chain or in the vicinity of the unstable periodic orbit corresponding to the resonance. The
difference of the quantum numbers  determines the excitation of the localized states which is
reveled using the zeros of the Husimi distribution. 
\end{abstract}
\maketitle

\noindent
{\bf Introduction:}
Hamiltonian classical systems have a variety of dynamical behaviors \cite{lichtemberg}. 
On one side are integrable systems with conserved quantities as 
degrees of freedom resulting in constrained dynamics around invariant tori. 
On the other extreme, chaotic systems are characterized by properties of mixing 
and ergodicity. The phase space is dynamically filled  
and only constrained by the conservation of the energy. 
The quantum mechanics of these dynamical systems have been intensively studied in the 
last forty years and the correspondence between classical and quantum mechanics has
been  established with solid grounds \cite{qchaos,chaosbook}.

It is unusual that a  generic system belongs to those extreme cases as
it would display {\it mixed dynamics}  where chaos coexist with regions of regular motion.   
The dynamics of mixed systems are more subtle mainly because regular and chaotic regions are 
connected by fractal boundaries. A standard way to understand  this complex dynamics  is
perturbing an integrable system. Its response to weak perturbations 
has been completely understood in terms of the celebrated Kolmogorov- Arnold-Moser (KAM) and the Poincar\`{e}-Birkhoff (PB) 
theorems \cite{KAMclasico,PB}.  The KAM theorem states that depending
on the rationality of the frequencies of the motion,  some of the invariant tori are deformed and survive  while others are 
destroyed. The consequence of the fate of rational tori is the survival of an equal even number of stable (elliptic) and 
unstable (hyperbolic) periodic orbits (PB theorem). In their vicinity of an stable orbit,  a chain of islands of regularity 
surrounded by a chaotic sea are developed \cite{lichtemberg}. This classical structures, usually called nonlinear resonance, 
have important influences in various phenomena from chemical systems, solid state physics to nano optics 
\cite{chem, geisel, optic}. 
  
Almost ubiquitous, the quantum mechanics of nearly regular and mixed systems is much less known than the integrable 
or chaotic cases. The relation between nonlinear resonances and avoided crossings (ACs) observed in the spectra of quantum systems  
has been a subject of several studies in the past  \cite{noid}. 
Using a semiclassical approach, nonlinear resonance were shown to be responsible for 
energy levels approach each other closely, exhibiting
avoided crossings,
instead of crossings as happen in integrable systems  \cite{uzer, ozorio}. 
More recently,  Brodier,  Schlagheck, and Ullmo \cite{RAT},  have developed a semiclassical theory of resonance assisted 
tunneling showing the coupling interaction term between states localized in invariant tori. Interesting enough,
a selection rule emerges because the interaction only occurs between states  with quantum number that differ in a {\it multiple} 
of the order of the resonance.  So, it is expected that the appearance of a nonlinear resonance is revealed in the quantum spectra with series of ACs of states localized in tori with quantum number that fulfil the selection rule.
In fact, this has been shown in Ref.  \cite{wisniacki}  when the multiple is equal to one.

In this letter we go one step forward and show such a systematic  for ACs between states with quantum 
number that differ in  a multiple greater that one. 
We compute the series of ACs  corresponding to a resonance of a paradigmatic system of quantum chaos studies 
and show that the states in the center of AC have a surpassing structure: one is an excited state localized the island
chain  and the other in the associated periodic orbit. 
This  shows that a classical
nonlinear resonance imprints clear signatures in the wave function morphologies.  
The eigenstates at the AC are carefully analyzed using the
Husimi distribution in phase space and their zeros \cite{voros}. The excitations of the localized structure are related
with the number of zeros in each island or in the vicinity of the unstable periodic orbit (PO) associate to the resonance. 

The systematic of ACs generated by a nonlinear resonance is a fertile field to test the building block  of the semiclassical theory of 
resonance assisted tunneling, that is, the  interaction coupling term between states localized in invariant tori \cite{RAT}.
For these reason,  we have studied the behavior of the series of ACs varying the value of the Planck constant $\hbar$. We clearly 
show an unexpected result that the semiclassical expression for the interaction coupling term works better
in the deep quantum regime seeing signs of an improvement of this theory seems necessary.

%
%
%
\vspace{0.5cm}
\noindent
{\bf The model:}
We study the Harper map in the unit square as a model system,
%
\begin{eqnarray}
  p_{n+1} & = & p_n + k \sin(2\pi q_n) \qquad ({\rm mod}\; 1), \nonumber \\
  q_{n+1} & = & q_n - k \sin(2\pi p_{n+1}) \qquad ({\rm mod}\; 1),
  \label{eq:1}
\end{eqnarray}
where $k$ is a parameter that measures the strength of the perturbation. 
This map can be understood as the stroboscopic version of the flow
corresponding to the (kicked) Hamiltonian
%
\begin{equation}
  H(p,q,t) = - \frac{k}{2\pi} \cos(2\pi p) - \frac{k}{2\pi} \cos(2\pi q)
    \sum_n \delta(t-n).
    \label{eq:2}
\end{equation}

\begin{figure}
\includegraphics[width=8.5cm]{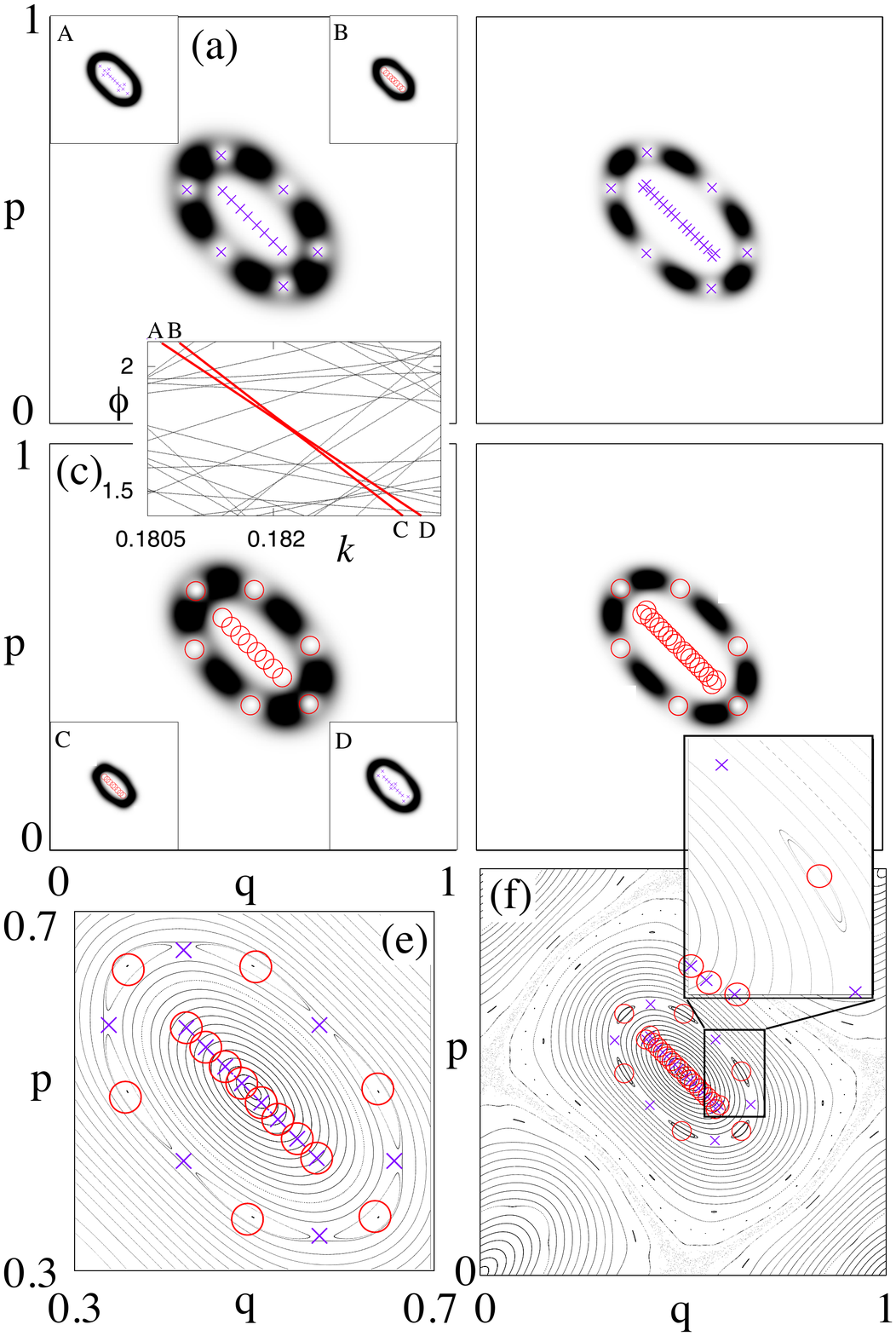}
 \caption{(Color online) (a) and (c) Husimi distribution of the states at the center of an AC obtained from the intersection of 
 quasistates  $9$ and $15$ for a Hilbert space with $N=160$. The zeros of the distribution are plotted with $\times$
 and $\bigcirc$.  In the inset (middle) of (a) and (c) the region of the spectra where the AC take place is plotted. The states before and after the AC are labeled A, B, C, D and are also plotted in the insets of (a) and (c). In (e) the zeros of the Husimi distributions (a) and (c) are plotted with the classical phase space at $k=0.1822$. 
(b) and (d) Husimi distribution of the states at the center of the AC obtained from the intersection of quasistates  $19$ and $25$ for $N=300$.  In (f) the zeros of the Husimi distributions (d) and (d) are plotted with a classical phase space at $k=0.1833$ In the inset of (d)-((f) a blow up of a part of panel (f) is plotted to see the location of the zeros of the Husimi distribution in more detail.  }
 \label{fig:1}
\end{figure}

\begin{figure}
\includegraphics[width=8.5cm]{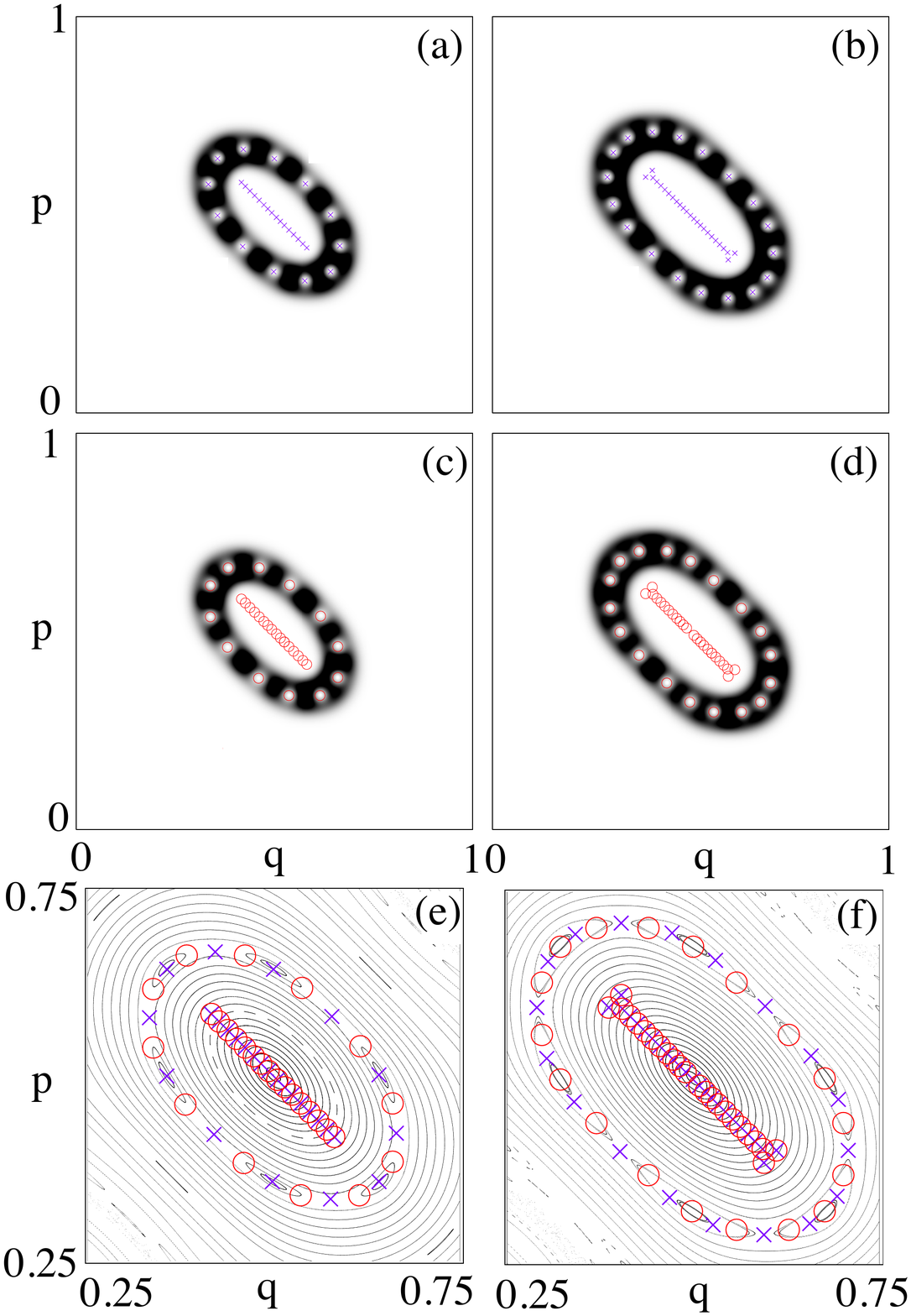}
 \caption{ (Color online)Husimi distribution of the eigenstates at the center of an AC obtained with the intersection of quasistates  $15$ and $27$ [(a) and (c)] and $26$ and $44$ [(b) and (d)] . The zeros of the distribution are also plotted with $\times$
 and $\bigcirc$.  In (e) and (f) the zeros of the Husimi distributions [(a), (c) and (b), (d)] are plotted with the corresponding part of the  classical phase space at $k=0.1809$. and $k=0.1978$. 
 }
 \label{fig:2}
\end{figure}

The Harper map comprises all the essential ingredients of mixed dynamics and is extremely simple from a numerical point of view.
For very small $k $, the dynamics described by the map is essentially regular, that is, the phase space is covered by invariant tori.  As $k$ gets bigger, 
non linear resonances (islands) start to appear following the KAM and PB theorems.  
The system presents a mixed dynamics with regions of regularity around the origin and the corners coexisting with chaos as 
shown in the bottom panels of Fig. \ref{fig:1} and \ref{fig:2}. 
For $k> 0.63$ there are no remaining visible regular islands due to  chaotic dynamics cover all the phase space.   

The quantum mechanics of the Harper map is described by the unitary time-evolution operator
\cite{Leboeuf1,Paz}
%
\begin{equation}
  \hat{U_k} = \exp[{\rm i}N k \cos(2\pi \hat{q})] \; \exp[{\rm i}N k \cos(2\pi \hat{p})],
  \label{eq:3}
\end{equation}
with $N=(2\pi\hbar)^{-1}$, that is, a Hilbert space of $N$ dimensions
for a fixed value of $\hbar$. This is due to the quantization on the torus which 
implies that the wave function should be periodic in both position and momentum 
representations. The semiclassical limit is reached as $N$ takes increasing values.

For a fixed value of $N$, the spectrum of eigenphases $\phi_i(k)$ and eigenfunctions 
$|\psi_i(k)\rangle$ of the evolution operator of the quantum map are obtaining by diagonalization of  Eq. \ref{eq:3}.  
The characteristics of  $|\psi_i(k)\rangle$ are analyzed using the Husimi distribution \cite{husimi}. 
The Husimi representation of an eigenstate of a quantum map is a quasiprobablity distribution in phase space that has 
exactly $N$ zeros in the unite  square \cite{voros}. 

Eigenphases $\phi_i(k)$ change linearly for very small strength of the perturbation $k$ 
and the  Husimi distribution  of the eigenstates $|\psi_i(k)\rangle$  are localized in the vicinity of invariant tori
\cite{Leboeuf1,voros}.   
Eigenstates with negative slope are centered in $(q,p)=(1/2,1/2)$  and  in $(q,p)=(0,0)$ for positive slope.  
A bigger absolute value of the slope implies that the state is nearer to the periodic point $(q,p)=(1/2,1/2)$ or 
$(q,p)=(0,0)$.
The states with maximun absolute value of the slope resemble a gaussian distribution centered in the mentioned periodic points
and corresponds to label $1$ (negative slope)  and $N$ (positive slope).  Exited states  has $n$ zeros inside the region of maximum probability, 
being $n+1$ its label for negative slope and $ (N-(n+1)$  for positive slope.


\vspace{0.5cm}
\noindent
{\bf Method:}
The influence of a non-linear resonance r:s  to quantum maps  is uncovered using the following numerical procedure.
 We consider a serie of eigenstates $|\phi_i(k)\rangle$ with $i=1,...i_{max}$ ($i_{max}< N/2$) for very small perturbation $k=k_0$.
and we associate for
each state $i$ a perturbed one with $k=k_0+\delta k$,  if the overlap $\langle \phi_i(k_0) |\phi_j(k_0+\delta k)\rangle$ is  the maximum of all $j$. Then, 
this procedure is repeated for perturbations $k=k_0+n \delta k$ with $n=2,...n_{max}$ an integer. Thus, we have associated a serie of perturbed states and eigenphases 
with the unperturbed one.  From now on, we refer quasistate $i$ to the serie of perturbed states associated to $|\phi_i(k_0)\rangle$ and quantum number to the label $i$ of the quasistate.  

If we join lines through the eigenphases of each quasistate, we can establish where two of the quasistates have  a crossing. In the vicinity of that intersection we find an AC of eigenstates 
that has the localized properties of the states at $k=k_0$.
If the dimensions $N$ of the Hilbert space is small enough, the previous procedure can be
done by visual inspection of the spectra as a function of the perturbation $k$.  
The value of  $\delta k$ is crucial for the success of the procedure: if it is small and for an $n$, $k=k=k_0+n \delta k$ falls close to an AC, the quasistate loses the localization properties 
related   with the unperturbed state and the method fails.   On the contrary, if  $\delta k$ is very large the method also fails  because
the phase of the quasistate has an erratic development and hence it is not possible to find its crossing with other quasistates.  

Once we have computed  the quasistates for a serie of unperturbed eigenstates $|\phi_i(k_0)\rangle$, we obtain the position of the corresponding ACs looking at the intersection of the eigenphases $i$ with $j$. If we are considering a non-linear resonance 
{\it r:s} with
$r$ the number of islands, we compute the series of ACs for quasistates with quantum numbers $i$ and  $j$ that  
\begin{equation}
\Delta n=|i-j|=lr
\end{equation}
with $l$ an integer. 

In Ref. \cite{wisniacki} the series of AC for two different nonlinear resonances of the Harper map were found for $l=1$ using
visual inspection of the spectra. This was feasible due to the small value of the dimension of the Hilbert space $N$. 
In the following, we show that using the described method it is possible to find the series of ACs for greater $l$ and $N$. 
  
\vspace{0.5cm}
\noindent
{\bf Results:}
The Harper map is a  mixed system that has an usual regular to chaotic transition. 
As the perturbation strength $k$ grows, non-linear resonances get bigger as
the surrounding chaotic layers and eventually disappear covered by the chaotic sea. 
As can be seen in the classical phase space showed in the bottom panels of Fig. \ref{fig:1} and \ref{fig:2}, the resonaces 6:1 
reach the largest size [see also Ref. \cite{wisniacki, RAT}].  Other resonances as  8:1, 10:1 and 14:1 take up an appreciable  
regions of phase space. 

Our main goal is to disentangle  the influence of non-linear resonance to the egenfunctions of a  mixed system. 
We focus  on the resonance 6:1 of the Harper map.
Using the method described below, we have found  the ACs associated to the intersections of the 
eigenphases of  quasistates with quantum number that differs 
in a multiple of 6, the order of the 
resonance 6:1, that is, for $\Delta n=6l$,
with  $l=1,2$ and $3$. 
The calculations are done for three sizes of the Hilbert space $N=80$, $160$ and $300$.
As an examples, in Fig. \ref{fig:1} and \ref{fig:2} we show
the morphologies of the eigenstates in these ACs.

In Fig. \ref{fig:1} the behavior of the wave functions in vicinity of an AC obtained from quasistates with quantum number that differ in 6 is exhibited.
Left panels corresponds to a Hilbert space with $N=160$ and right panels to $N=300$.
In the insets of the left panels of Fig. \ref{fig:1}  we can see the Husimi distributions of  eigenfunctions before and after the AC 
that was obtained from the intersection of quasistates 9 and 15. 
In the central inset of Fig. \ref{fig:1} (a) and (c)  we show the region of the spectra where the AC take place. The eigenphases
of the states that have the AC are plotted in red lines. As it is usual,  the states exchange their distributions 
upon AC  [$(A) \leftrightarrow (D)$ and $(B) \leftrightarrow (C)$]. 
But surprising enough, the distributions at the center of the AC are highly localized.
One state has the maximum of the probability of the Husimi distribution in the vicinity of the island chain and have 6 zeros 
near the corresponding unstable PO [see Fig. \ref{fig:1}(a)]. The other state [Fig. \ref{fig:1}(c)], is localized in the
vicinity of  the unstable PO and one zero in the center of each island is observed. 
This is better displayed in Fig. \ref{fig:1}(e) where a part of the classical phase space and  the zeros of the Husimi distribution
of states of Fig. \ref{fig:1} (a) and (c) are shown. The zeros of the spates (a) and (b) are plotted with $\times$
and with $\bigcirc$ for (c) and (d).
In the right panels of Fig. \ref{fig:1}, we show a similar example for $N=300$ and quasistates 19 and 25.
A blow up of the vicinity of an islands of the resonance $6:1$ is shown in the inset in the middle of  panels (d) and (f) of Fig.
\ref{fig:1}.

When the AC is between states with quantum number that differ in a multiple (greater that one) of the number 
of islands in the chain, the morphologies of the eigenstates  are more impressive. 
As an example of  this phenomenon,  in Fig. \ref{fig:2} we show  the Husimi distributions  at the center of AC between states with quantum numbers that
 differ in 12 [Fig. \ref{fig:2}(a),(c) and (e)] and 18 [Fig. \ref{fig:2}(b), (d) and (f)]. 
 As can be seen in the cases
 displayed in Fig. \ref{fig:1}, one of the states is localized in the island chain, but in Fig. \ref{fig:2}(a) there is one zero inside
 each island and two zeros for Fig. \ref{fig:2}(b). This fact point out that these states are excited states of the island chain. 
  The other states, 
 Fig. \ref{fig:2}(c) and (d) are localized in the corresponding unstable PO and the zeros  are accumulated inside the islands.
 This fact resemble the behavior of scar functions builded for chaotic system \cite{vergini,wisniacki2}.
 To see the location of the zeros in more detail, in Fig.  \ref{fig:2} (e) and (f) we plot a part of the classical phase space and the zeros of the Husimi distributions of panels 
(a)-(c) and (b)-(d).
We have seen that when  $\Delta n=24$ the structure of the wave functions at the AC have the same systematic with one more 
zero in each island of the chain. 
In summary, in the center of an AC with $\Delta n=6 l$, one of the states is localized in the vicinity of the island chain and has
$l-1$ zeros in each island, whereas the other state has the maximum probability around the unstable PO of the resonance
and has $l$ zeros in each island.  
It is important to note that in Figs. \ref{fig:1} and \ref{fig:2} only the zeros of the Husimi distribution that are inside or over
the regions of maximum probability are plotted.  Other zeros that lie outside these regions and have exponential small
probability are not displayed.

The semiclassical  theory  of resonance assisted tunneling predicts  a coupling strength between quasi modes located on opposite sides of a nonlinear resonance and therefore an eigenphases difference $\Delta \phi$ for the ACs considered before.
This theory  was recently 
developed and applied in several situations \cite{RAT,RAT2,RAT3} . 
The starting point  is the classical secular perturbation 
theory  which allows to construct an effective time independent Hamiltonian that describes  the local dynamics near a r:s  resonance of the map,
\begin{equation}
H_{r:s} \simeq H_0(I_{r:s})+\sum_{l=1}^{\infty} V_{r,l}(I_{r:s})\cos(l r \theta+\phi_l)
\end{equation}
with $H_0(I_{r:s})$ an integrable approximation of the Hamiltonian of the map \cite{RAT}.
This effective Hamiltonian entails a selection rule that an eigenstate of the unperturbed Hamiltonian of a quantum number $n$ 
can be coupled to another state of a quantum number $n + lr$ (l an integer) with a strength proportional to $V_{r,l}$.

 \begin{equation}
  V_{r,l} e^{i \varphi_l} = \frac{1}{i\pi rs \tau}
     \int_0^{2\pi} \exp(-i r l \theta) \; \delta I_{r:s}(\theta) d\theta
  \label{eq:5}
\end{equation}
$\delta I_{r:s}(\theta)$ being given by
%
\begin{equation}
   \delta I_{r:s}(\theta) = I^{(-1)}(I_{r:s},\theta)-I_{r:s},
  \label{eq:6}
\end{equation}
where $I^{(-1)}(I_{r:s},\theta)$ is  the action variable obtained by applying the inverse 
Poincar\'e map to $(I,\theta)$. 
The interaction coupling strength [Eq. \ref{eq:5}] is numerically computed  
following the next setps. 
First, the resonant periodic torus is found, and its action $I_{r:s}$ calculated.
Then, it is computed a large number of points $(q_i,p_i)$ and its corresponding angle variable $\theta_i$ belonging 
to the resonant tours . 
We applied a 
back-propagation for these points with the exact inverse map.
The associated perturbed action of each point $I^{(-1)}(I_{r:s},\theta_i)$ is computed by numerical propagation in a 
complete cycle. Using these quantities the coupling interaction $V_{r,l}$ is calculated with Eq. \ref{eq:5}. 
Finally,  a semiclassical approximation of  the eigenphases differences $\Delta \phi$ produced by a resonace $r:s$
for AC that comes from the intersection of quasistates eigenphases with $\Delta n=l r$ results
\begin{equation}
 \Delta \phi \approx |V_{r,l}|/2 \hbar. 
 \label{eq:8}
\end{equation}
  
\begin{figure}
\includegraphics[width=8.5cm]{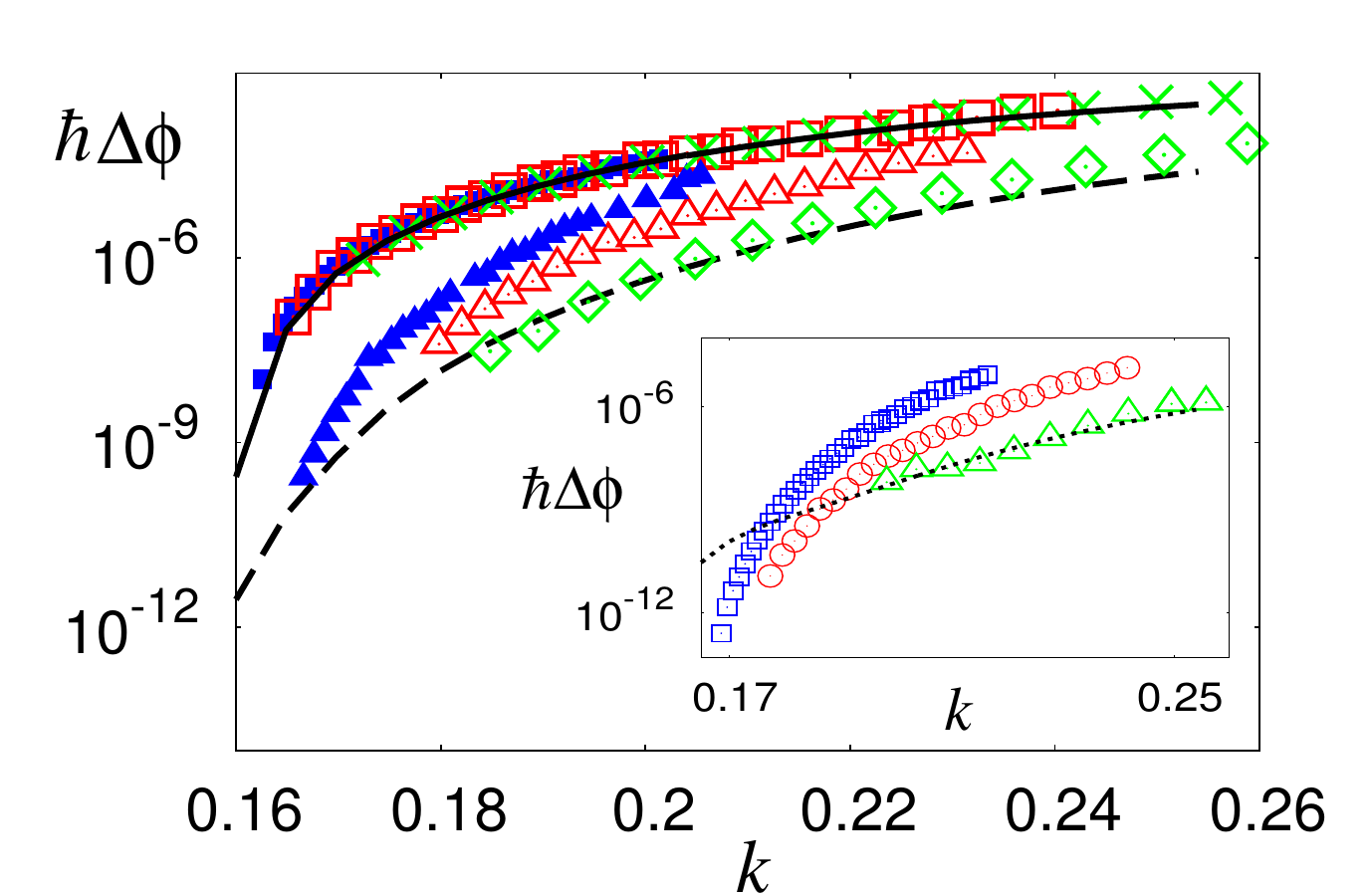}
 \caption{(Color online:) Eigenphase difference $\Delta \phi$ (scaled with $\hbar$) as a function of the perturbation strength $k$  for  ACs associated with
 the nonlinear resonance 6:1 of the Harper map. Symbols corresponds to ACs obtained from the quantum spectra.
 For $\Delta n=6$ it is used  $\times$ for $N=80$,   
 $\Box$ for $N=160$ and  $\blacksquare$ for $N=300$. $\Delta n=12$ is plotted with 
 $\Diamond$ for $N=80$,   $\bigtriangleup$ for $N=160$ and $\blacktriangle$ for $N=300$. In the inset $\Delta \phi$ corresponds to ACs with $\Delta n=18$ with 
 $\bigtriangleup$ for $N=80$,   $\bigcirc$ for $N=160$ and  $\square$ for $N=300$.
 The semiclassical  prediction  of the eigenphase difference Eq. \ref{eq:8} is plotted with lines: 
 $m=1$ solid,  $m=2$ dashed and  $m=3$ dotted (in the inset). 
}
 \label{fig:3}
\end{figure}

We have computed the semiclassical 
approximation of the eigenphase difference for the resonance 6:1 with $l=1,2$ and $3$ as a function of the 
strength $k$. The integral was done using the 7-point Newton-Cotes formula and with an integrable approximation of the Harper 
Hamiltonian  [Eq.~(\ref{eq:2})] up to 5th order that was obtained using the Baker-Campbell-Hausdorff  formula
in Eq. \ref{eq:3} and the semiclassical relation between quantum commutator and the Poisson brackets \cite{bch}.
In Fig. \ref{fig:3}, the semiclassical approximation of eigenphase difference $\Delta \phi$
is plotted with lines ($l=1$ solid, $l=2$ dashed and $l=3$ dotted). 
The eigenphases differences $\Delta \phi$ for the ACs between states with $\Delta n=6l$, with  $l=1,2$ and $3$  are 
plotted with symbols. These
eigenphases differences were  obtained from the quantum spectra using the method presented before for three 
values of the number of states of the Hilbert space $N=80$, $160$ and $300$. 
 
\begin{figure}
\includegraphics[width=8.5cm]{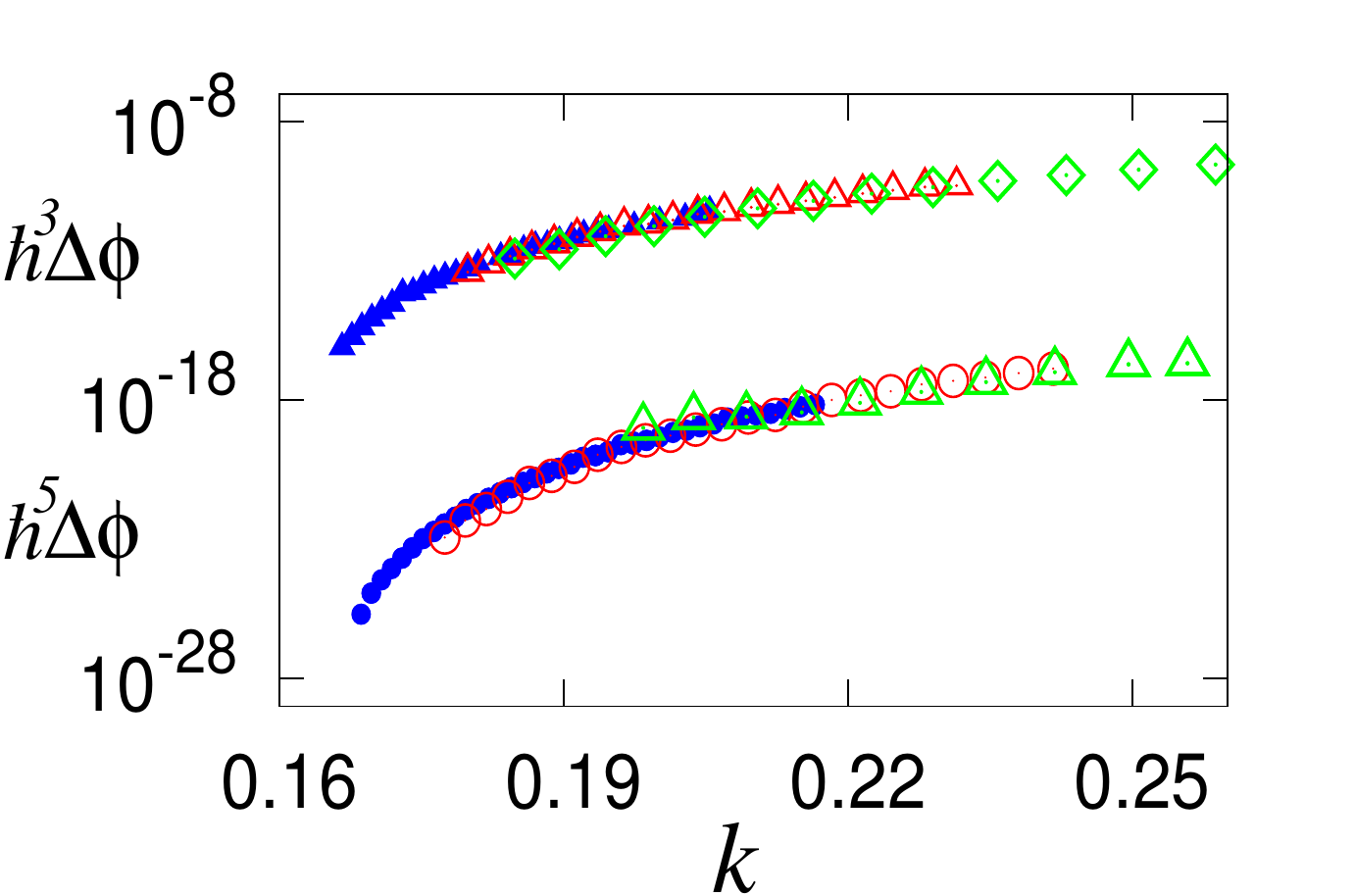}
 \caption{(Color online:) Scaled eigenphase difference $\Delta \phi$ as a function of the perturbation strength $k$  for  ACs 
 with $\Delta n=12$ and $\Delta n=18$. The scaling factor is  $\hbar^3$ for
 $\Delta n=12$ and $\hbar^5$ for $\Delta n=18$.  Symbols are the same as Fig. \ref{fig:3} }
 \label{fig:4}
\end{figure}

 In Fig. \ref{fig:3} we can see
that, contrary to the expectations, the semiclassical approximation works very nice for all $l$ only in the case of $N=80$, the minimum value of the considered number of states of the Hilbert space. 
This fact indicates that semiclassical expression of $\Delta \phi$ [Eq. \ref{eq:8}] works in the deep quantum regime and
separates from quantum results as $N$ increase. The discrepancies for large $N$ is becoming
an strong evidence that the semiclassical theory of the interaction coupling needs an improvement
\cite{RAT}. We note that similar behavior was observed  in the tunneling-induced level splittings of a very simple one dimensional model computed with the semiclassical theory of resonance assisted tunneling\cite{RAT3}.
In Fig. \ref{fig:4} we show an unexpected scaling of the $\Delta \phi$ for $l=2$ and $l=3$. 
Eigenphase difference scales with $\hbar^3$ for $l=2$
and with $\hbar^5$ for $l=3$. This could indicate the existence of an effective $\hbar$ that depends on $l$  and
could be a clue in the improvement of the semiclassical theory of interaction coupling strength [Eq. \ref{eq:5}] .

\vspace{0.5cm}
\noindent
{\bf Final Remarks:}
In this letter we have shown that a classical non linear resonance imprints a systematic influence in the quantum eigenvalues
and eigenfunctions of a mixed system.   We have found an universal structure embedded in the spectra: 
states localized in tori interact in AC if the quantum numbers differ in a multiple of the order of the resonance. These series
of AC are observed when a parameter of the system is varied producing the development of the resonance characterized by 
a chain of islands.  Surprisingly, eigenstates in the middle of the AC has a particular morphology.
One state is localized in the vicinity of the unstable PO associated to the resonance. The other state is localized on the island chain. 
The difference of the quantum numbers of the unperturbed states that are localized in tori and interact in the AC determines
the distribution of the zeros of the Husimi function of the states.
These findings could be of importance in the design of optical micro cavities\cite{RATexp,wiersig,martina}. In those devices it is desirable to obtain a specific directional 
light emission that could be accomplished tuning a system parameter to reach to an AC where the states have a desirable localization.  
 
We have compared the eigenphases gaps of the AC with a semiclassical prediction based on the theory of resonance assisted tunneling \cite{RAT}.  We have shown that the semiclassical prediction deviate from the quantum results as we reach the semiclassical limit. This unexpected result indicate  that an improvement of the this theory is needed.
 
\acknowledgments
We thank  Marcos Saraceno for stimulating discussions.  Financial support from  from ANCyPT (PICT 2010-1556), UBACyT,
and CONICET is acknowledged.

\end{document}